\documentclass[reprint,superscriptaddress,amsmath,amssymb,pra,longbibliography]{revtex4-2}

\usepackage{mathtools}
\usepackage{graphicx}
\usepackage[colorlinks=true,linkcolor=blue,citecolor=blue,urlcolor=blue,pdfauthor={ },pdftitle={ },pdfsubject={ },pdfkeywords={ }]{hyperref}
\usepackage{times}
\usepackage[qm]{qcircuit}
\usepackage{braket}
\usepackage{enumitem}
\usepackage{dcolumn}
\usepackage{bm}

\usepackage{siunitx}
\usepackage{upgreek}
\usepackage{standalone}

\usepackage{color}
\usepackage{tikz}
\usepackage{pgf}

\usepackage{hyperref} 
\hypersetup{
	pdfborderstyle={/S/U/W 1},
	colorlinks=true,
	linkcolor=blue,
	urlcolor=blue,
	citecolor=blue,
	filecolor=blue
}

\makeatletter
\DeclareRobustCommand{\mymediumseries}{%
	\not@math@alphabet\mymediumseries\relax
	\fontseries{m}\selectfont
}
\makeatother

\begin{document}
	
\title{Investigating Lipkin–Meshkov–Glick Model and Criticality-Enhanced Metrology \\
in a Coherent Ising Machine}%

\author{Shuang-Quan Ma}
\affiliation{School of Physics and Astronomy, Applied Optics Beijing Area Major Laboratory, Beijing Normal University, Beijing 100875, China}
\affiliation{Key Laboratory of Multiscale Spin Physics, Ministry of Education, Beijing Normal University, Beijing 100875, China}

\author{Jing-Yi-Ran Jin}
\affiliation{School of Physics and Astronomy, Applied Optics Beijing Area Major Laboratory, Beijing Normal University, Beijing 100875, China}
\affiliation{Key Laboratory of Multiscale Spin Physics, Ministry of Education, Beijing Normal University, Beijing 100875, China}

\author{Chen-Rui Fan}
\affiliation{School of Artificial Intelligence, Beijing Normal University, Beijing, 100875, China}

\author{Chuan Wang}
\affiliation{School of Artificial Intelligence, Beijing Normal University, Beijing, 100875, China}

\author{Qing Ai}
\email{Electronic mail: aiqing@bnu.edu.cn}
\affiliation{School of Physics and Astronomy, Applied Optics Beijing Area Major Laboratory, Beijing Normal University, Beijing 100875, China}
\affiliation{Key Laboratory of Multiscale Spin Physics, Ministry of Education, Beijing Normal University, Beijing 100875, China}

\date{\today}%
\begin{abstract}
Quantum criticality has received extensive attention due to its ability to significantly enhance quantum sensing. But its realization and control in many-body quantum systems remain challenging. We present an effective scheme to simulate the Lipkin–Meshkov–Glick (LMG) model using a coherent Ising machine (CIM) composed of a network of degenerate optical parametric oscillators (DOPO). In our work, the spin variables of the LMG model are mapped onto the phases of DOPO pulses, and the spin-spin interactions are realized by all-to-all couplings among them. Through our investigation of the critical behavior in the antiferromagnetically coupled LMG model in the thermodynamic limit, i.e., $N\rightarrow\infty$, and its application in quantum sensing near the critical point, we verify that the CIM does not only effectively capture the second-order quantum phase transition (QPT) at the critical point but also reconstructs its complete phase diagram under ferromagnetic coupling. Furthermore, we demonstrate how the critical dynamics of this simulation platform can be utilized for quantum-enhanced metrology, achieving a measurement precision that diverges near the critical point of the LMG model. These results highlight the capability of the CIM as a flexible experimental platform for investigating the QPT in the fundamental quantum magnetic models, providing valuable insights into quantum simulation and critical phenomena.
\end{abstract}
\maketitle

\section{Introduction}
Quantum phase transition (QPT)~\cite{Sachdev2011,science1201607,NatureJulian,PhysRevX031028} is the phase transition at the zero temperature characterized by non-analytic changes in the properties of the ground state as a control parameter is tuned and exists in various fields, including quantum optics~\cite{PhysRevX011016,PhysRevX011012,Noh2017}, condensed-matter physics~\cite{PhysRevLett247402,NP2018,Hartmann2016}, and quantum-information science~\cite{PhysRevA032110,Scaling2002,Guifre2003,PRXQuantum020337,PRXQuantum020342}. In particular, the non-analyticities in the ground-state energy~\cite{RevModPhys315,Vojta2003} and the enhanced susceptibility near the critical point make the QPT a powerful resource for quantum-enhanced sensing~\cite{Degen2017,Kwon2019,di2023critical,PhysRevA052621,PhysRevLett040801,Alushi2025,Frerot2018,Yu2025,Sarkar2025,Agarwal2025}. Significant experimental milestones in criticality-enhanced sensing have been achieved using nuclear magnetic resonance~\cite{liu2021} and Rydberg atoms~\cite{ding2022}. The achieved measurement precision surpasses the standard quantum limit $M^{-1/2}$~\cite{Giovannetti}, where $M$ is the number of measurements. Recent theoretical investigations have extended criticality-enhanced sensing to various platforms, proposing protocols based on the quantum Rabi model~\cite{Zhu2023,Cai2021,Ying2022,Garbe2020,Chen2025}, the parametrically-driven Jaynes-Cummings model~\cite{lu2022}, the parametrically-driven Tavis-Cummings model~\cite{Mattes2025,Lu23}, and the cavity-Ising chain coupling model~\cite{Gammelmark2011}, all aiming to demonstrate the potential of phase-transition points across different physical platforms for quantum metrology. { Additionally, a driven-dissipative quantum sensor is proposed and analyzed~\cite{Ilias2024npjQI,Ilias2022prx,Yang2023prx}. By simultaneously harnessing dissipative criticality and an efficient continuous-readout mechanism, the sensor achieves high-precision detection of oscillating electric-field gradients, with its accuracy surpassing the standard quantum limit. The scheme has also been demonstrated to be robust against practical experimental imperfections.} The Lipkin-Meshkov-Glick (LMG) model stands as a typical example for studying the QPT. It features infinite-range spin-spin interactions, and is exactly solvable, and exhibits a second-order QPT from a symmetry-broken ferromagnetic phase to a polarized paramagnetic phase~\cite{Dusuel2004,Dusuel2005,Kwok032103,Latorre2005,Salvatori2014,lee2014,Orus2008,Morrison2008,Ribeiro2007,Debecker140403,Debecker2025}. Originally proposed in nuclear physics to describe shape phase transitions in atomic nuclei~\cite{Lipkin1965}, the LMG model has also been extensively studied in the quantum information processing~\cite{Ma2009,Vidal,Vidal2004,Ribeiro2008,xu2020,Barthel2006,MA201189}. Importantly, its critical region features spin squeezing and multipartite entanglement that scale favorably with system's size, offering a theoretical pathway toward achieving the Heisenberg limit $M^{-1}$~\cite{Holland}. {However, the experimental realization of the LMG model's criticality remains highly challenging. A recent experiment has reported macroscopic self-trapping and a dynamical phase transition in a momentum-space Bose-Einstein condensates double-well platform~\cite{Macroscopic2025arXiv}.} The realization of macroscopic all-to-all coupled spin ensembles requires precise control of collective quantum states, which is hindered by decoherence and scalability in the traditional solid-state platforms, such as cavity quantum electrodynamics~\cite{Morrison2008,Zhang2017,Liu013601,Mivehvar02012021,Takano033601}, Rydberg ensembles~\cite{Hines2023prl,Ding2022NPhys} and trapped-ion systems~\cite{Britton2012Nature}. This challenge motivates the pursuit of quantum simulation~\cite{Georgescu2014,Peng2025QST}, which employs controllable quantum systems to simulate complex ones.

Over the past two decades, significant progress has been made in the quantum simulation using  experimental platforms such as ultracold atoms~\cite{Makhalov120601}, trapped ions~\cite{Unanyan2003}, and solid-state spins~\cite{zhou2017,you2014,zou2014}. However, these platforms still face challenges in scalability, connectivity, and controllability. For instance, ultracold atoms offer large system's size but are typically limited to short-range interactions~\cite{Blochs885,Sciencel3837}. Trapped ions and superconducting qubits can achieve long-range couplings but are generally confined to smaller scales~\cite{blatt2012,barreiro2011,Science1208001,monroe2021,Lv021027}.

In this context, the coherent Ising machine (CIM) provides a uniquely-scalable and controllable platform~\cite{wang2013}. Based on networks of degenerate optical parametric oscillators (DOPO), the CIM encodes spin variables in the phase states of individual DOPO and implement all-to-all couplings via optical injection~\cite{Science5178,Science4243,SA100000,babaeian2019}. Such systems have realized fully-connected networks with up to 100,000 nodes~\cite{SA100000}, far exceeding the connectivity achievable in the solid-state systems. Moreover, the intrinsic dissipative nonlinear dynamics of the DOPO naturally support both equilibrium QPTs and the non-equilibrium critical phenomena, while the homodyne detection enables quantum-nondemolition readout of the order parameters~\cite{Woo1971,gatti1995,drummond2002,Roy21,drummond2005,Science1086489,roy2021,Goto4065825,Bello083901}.

Although prior research on the CIM has primarily focused on solving combinatorial optimization problems~\cite{bohm2019,yamamoto2017,Lu2023OE,Fan2025OE,Li2025QST,rah2023,yamamura2024,SA2372,SA0823,SA8080,SA7223}, their potential for simulating equilibrium QPT and exploiting criticality for quantum sensing remains largely unexplored. In this work, we develop a CIM-based quantum simulation framework that does not only reproduce the QPT of the LMG model but also can utilize its criticality for quantum-enhanced sensing. Recently, Cai {\it et al.} proposed a criticality-enhanced quantum-sensing protocol~\cite{Cai2021}, which predicts that the quantum Fisher information (QFI) will diverge near the critical point. Based on this protocol, our numerical results demonstrate that the QFI for both the LMG model and the CIM indeed exhibits divergent behavior at the critical point. Meanwhile, the measurement precision for the simulated quadrature component, i.e., the inverted variance, is also significantly enhanced near the critical point, further confirming the effectiveness of the critical effect in improving sensing sensitivity. This approach bridges quantum optics and many-body physics, demonstrating the CIM as a scalable simulator for the QPT and providing a blueprint for probing singular quantum phases in programmable photonic lattices.

This paper is structured as follows. In Sec.~\ref{sec:LMG}, we introduce the LMG model as well as the relevant content of the QPT and outline the concept based on the metrics of quantum criticality. In Sec.~\ref{sec:CIM}, we employ the CIM to simulate the QPT and criticality for quantum-enhanced sensing of the LMG model. In Sec.~\ref{sec:Results}, we mainly analyze and discuss the results of the CIM simulating the LMG model and reproducing its QPT and critical-enhancement quantum sensing capabilities. Section~\ref{sec:Conclusions} concludes this paper.

\section{Lipkin-Meshkov-Glick Model and Quantum Criticality}\label{sec:LMG}

In this section, we begin with a theoretical analysis of the QPT in the LMG model in Sec.~\ref{sec:qpt_lmg}. We derive key analytical results that characterize the critical behavior of the model. Building on this foundation, Sec.~\ref{sec:qm_lmg} extends the theoretical framework to quantum metrology, examining how the sensitivity of parameter estimation can be enhanced near the critical point of the QPT. Together, these theoretical explorations set the stage for the subsequent numerical simulation of the LMG model using the CIM, with a focus on probing quantum critical phenomena and critically-enhanced quantum metrology.

\subsection{Quantum Phase Transition in the LMG Model}\label{sec:qpt_lmg}

We begin with the LMG model, an important paradigm in quantum optics. The LMG model describes a set of $N$ spin-1/2 particles interacting via an anisotropic XY-type Hamiltonian and coupled to an external transverse magnetic field $h$, exhibiting a second-order QPT. {Hereinafter, we set $\hbar = 1$.} The Hamiltonian associated with the LMG model can be expressed as \cite{Lipkin1965,Ribeiro2007,Debecker140403,Debecker2025}
\begin{equation}
	H=\frac{\lambda}{N}\sum_{1\leq i<j\leq N}\left(\sigma_{i}^{x}\sigma_{j}^{x}-\sigma_{i}^{y}\sigma_{j}^{y}\right)+h\sum_{j=1}^{N}\sigma_{j}^{z},
\end{equation}
where $\sigma_{j}^{(\alpha)}(\alpha=x,y,z)$ are Pauli operators of the $j$th qubit and $\lambda$ is the coupling strength. The prefactor $1/N$ is necessary to obtain a finite free energy per spin in the thermodynamic limit $N\rightarrow\infty$. Here we focus on the antiferromagnetic case, i.e., $\lambda>0$, and without loss of generality, we set $\lambda = 1$.

The collective nature of the model allows us to write the Hamiltonian in terms of the total spin operators, i.e., $S_{\alpha}=\sum_{j=1}^{N}\sigma_{j}^{\alpha}/2$, such that
\begin{equation}
H_{\rm anti}=\frac{2\lambda}{N}\left(S_{x}^{2}-S_{y}^{2}\right)+2h S_{z},
\label{eq2}
\end{equation}
where a constant energy shift has been neglected. The phase diagram consists of two distinct regions and exhibits a second-order QPT when $h \!\!=\!\!\lambda$ \cite{Vidal2004,Dusuel2005,Ribeiro2007}. The phase at $h \!\!>\!\!\lambda$ is called the normal phase, while the symmetry breaking phase at $h \!\!<\!\!\lambda$ is called the deformed phase. In the limit of weak interaction, the LMG model can be solved exactly by mapping it to $N$ bosons in a double well, while in the thermodynamic limit $N \!\!\rightarrow\!\!\infty$, it can be solved through the Holstein-Primakoff transformation \cite{HP1940,Dusuel2004,Dusuel2005,Kwok032103}. The latter approach is also a good approximation for $N \!\!\gg\!\! 1$ and low excitation~\cite{Hirsch2013}. In the Holstein-Primakoff transformation~\cite{HP1940}, we define
\begin{subequations}\label{eq:3}
\begin{align}
S_+ &= S_x + iS_y = \sqrt{N} \sqrt{1 - \frac{a^\dagger a}{N}} a^\dagger, \label{eq:3a} \\
S_- &= S_x - iS_y = \sqrt{N} \sqrt{1 - \frac{a^\dagger a}{N}} a, \label{eq:3b} \\
S_z &= a^\dagger a - \frac{N}{2}. \label{eq:3c}
\end{align}
\end{subequations}
with $a$ and $a^\dagger$ the annihilation and creation operator of bosonic mode respectively, satisfying $[a,a^\dagger]=1$. Under the condition that $N \gg1$ and low excitation, i.e., $a^{\dagger}a\ll N$, we have \cite{HP1940}
\begin{subequations}\label{eq:4}
\begin{align}
S_{+}&\approx\sqrt{N}a^{\dagger},\label{eq:4a} \\
S_{-}&\approx\sqrt{N}a,\label{eq:4b} \\
S_{z}&=a^{\dagger}a-\frac{N}{2}. \label{eq:4c}
\end{align}
\end{subequations}
Substituting above formulas into Eq.~(\ref{eq2}) yields
\begin{align}
\tilde{H}_{\rm anti}&\approx 2ha^{\dagger}a+\lambda\left(a^{\dagger 2}+a^{2}\right)-hN.
\label{eq:Hanti}
\end{align}
By Bogoliubov transformation \cite{Sachdev2011}
\begin{align}
b&=a\cosh r+a^{\dagger}\sinh r,
\end{align}
with $\tanh{(2r)}=-\lambda/h$, and this relation holds only when $h > \lambda$. We can diagonalize the Hamiltonian as
\begin{align}
\tilde{H}_{\rm anti} =E_k^{\rm anti} b^\dagger b +E_g^{\rm anti},
\end{align}
where the eigen energies and the ground-state energy are respectively
\begin{align}
	E_k^{\rm anti}&=2\sqrt{h^2-\lambda^2}, \\
	E_g^{\rm anti}&=\sqrt{h^2-\lambda^2}-h(N+1).
\end{align}

In the ferromagnetic phase, i.e., $ h<\lambda$, the spontaneous magnetization of the ground state is aligned in the $xy$-plane, e.g. along the $x$-axis. 
By defining
\begin{subequations}\label{eq:10}
\begin{align}
\tilde{S}_{x} &= S_{x}, \label{eq:10a}\\
\tilde{S}_{y} &= S_{y} \cos\theta + S_{z} \sin\theta, \label{eq:10b} \\
\tilde{S}_{z} &= -S_{y} \sin\theta + S_{z} \cos\theta,	\label{eq:10c}
\end{align}
\end{subequations}
with $\cos\theta=h/\lambda$, the Hamiltonian (\ref{eq2}) is rewritten as
\begin{align}
H_{\rm fer} =& \frac{2\lambda}{N} \tilde{S}_{x}^{2} - \frac{2\lambda}{N} \left( \frac{h}{\lambda} \tilde{S}_{y} - \sqrt{1 - \frac{h^2}{\lambda^2}} \tilde{S}_{z} \right)^2 \nonumber \\
&+2h \sqrt{1 - \frac{h^2}{\lambda^2}} \tilde{S}_{y} + \frac{2h^2}{\lambda} \tilde{S}_{z}.
\end{align}
Furthermore, applying the Holstein-Primakoff transformation, 
\begin{subequations}\label{eq:12}
\begin{align} 
	\tilde{S}_{z} &= a^{\dagger}a-\dfrac{N}{2}, \label{eq:12a} \\ 
	\tilde{S}_{+} &= \tilde{S}_{x}+ i\tilde{S}_{y} \simeq \sqrt{N}a^{\dagger}, \label{eq:12b} \\ 
	\tilde{S}_{-} &= \tilde{S}_{x}- i\tilde{S}_{y} \simeq \sqrt{N}a, \label{eq:12c}
\end{align}
\end{subequations}
in the thermodynamic limit and low excitation, i.e., $N \!\!\rightarrow\infty$ and $a^{\dagger}a\ll N$, we have
\begin{align} 
H_{\rm fer} =& \left( 3\lambda - \frac{h^2}{\lambda} \right) a^{\dagger}a + \left( \frac{\lambda}{2} + \frac{h^2}{2\lambda} \right) \left( a^{\dagger 2} + a^2 \right) \nonumber \\
&-\frac{\lambda N}{2} - \frac{h^2 N}{2\lambda}.
\end{align}
Using Bogoliubov transformation with $\tanh{(2r)}=(\lambda^2+h^2)/(3\lambda^2-h^2)$, we can diagonalize the Hamiltonian as
\begin{align}
	H_{\rm fer} =E_k^{\rm fer} b^\dagger b +E_g^{\rm fer},
\end{align}
with the eigen energies and the ground-state energy being respectively
\begin{align}
	E_k^{\rm fer}&=2\sqrt{2}\sqrt{\lambda^2-h^2}, \label{eq:15a} \\
	E_g^{\rm fer}&=\frac{E_k^{\rm fer}-3\lambda}{2} + \frac{h^2}{2\lambda}-(\frac{\lambda}{2} + \frac{h^2}{2\lambda})N. \label{eq:15b}
\end{align}
To summarize, the ground-state energy is
\begin{align}
	E_g = 
	\begin{cases} 
		 \sqrt{h^2-\lambda^2}-h(N+1), & h \geq \lambda, \\
		 \sqrt{2}\sqrt{\lambda^2-h^2}-\frac{3\lambda^2-h^2}{2}-(\frac{\lambda}{2} + \frac{h^2 }{2\lambda})N, & h < \lambda.
	\end{cases}
\end{align}
Thus, the rescaled ground-state energy is
\begin{align}
	e_g \equiv \frac{E_g}{N}= 
	\begin{cases} 
		-h, & h \geq \lambda, \\
		-\frac{\lambda}{2} - \frac{h^2}{2\lambda}, & h < \lambda,
	\end{cases}
\end{align}
which clearly shows the non-analyticity at the quantum critical point, i.e., $h = \lambda$.

\subsection{Quantum Metrology}\label{sec:qm_lmg}

To study critically-enhanced quantum sensing in the LMG model, we neglect the constant term in Hamiltonian~(\ref{eq:Hanti}) to obtain the required Hamiltonian
\begin{align}
	\tilde{H}_{\rm anti}&\approx 2ha^{\dagger}a+\lambda\left(a^{\dagger 2}+a^{2}\right).
	\label{eq:HLMG}
\end{align}
For simplicity, we define $\omega=2h$ and $g=\lambda/h$, so that Eq.~(\ref{eq:HLMG}) becomes $\tilde{H}_{\rm anti}=\omega[a^{\dagger}a+g/2(a^{\dagger 2}+a^{2})]$. We define the quadrature operators as 
\begin{eqnarray}
	X=\frac{a+a^\dagger}{\sqrt{2}},\quad
	P=\frac{a-a^\dagger}{\sqrt{2}i}, 
	\label{eq:XY}
\end{eqnarray} \\
which satisfy the commutation relation $[X,P]=i$. The Hamiltonian can be written as
\begin{align}
	\tilde{H}_{\rm anti}=\frac{\omega(1-g)}{2}\left[P^2+\left(1+\frac{2g}{1-g}\right)X^2\right].
\end{align}
To analyze the QFI for parameter estimation near the critical point, we follow the approach in Ref.~\cite{Pang022117}. We set $H_\eta=H_0+ \eta H_1$, where this Hamiltonian satisfies
\begin{align}
	[H_\eta,\zeta]=\sqrt{\Lambda}\zeta,
\end{align}
with $\zeta=i\sqrt{\Lambda}C-D$, $C=-i[H_0,H_1]$, and $D=[iC,\tilde{H}_{\rm anti}]$. Let $H_0=\omega(1\!\!-\!\! g) P^{2}/2$ and $H_1=\omega(1\!\!-\!\! g) X^{2}/2$, so that $\eta=1+2g/(1-g)$. The QFI for the estimation of the parameter $\eta$ around the critical point can be expressed as~\cite{Cai2021}
\begin{equation}
	I_{\eta} \simeq \frac{4[\sin(\sqrt{\Lambda}t)-\sqrt{\Lambda}t]^{2}}{\Lambda^{3}}\mathrm{Var}[D]_{|\varphi\rangle},
\end{equation}
where $\Lambda=4\omega^2(1-g^2)$ and $|\varphi\rangle$ is the initial state of the bosonic field. Applying this analysis to our parametrically-driven bosonic system, the QFI for the parameter $g$ is
\begin{equation}
	I_{g}(t)\simeq16(1+g)^2\frac{[\sin(\sqrt{\Lambda_g}\omega t)-\sqrt{\Lambda_g}\omega t]^{2}}{\Lambda_g^{3}}\mathrm{Var}[X^2]_{|\varphi\rangle},
	\label{eq6}
\end{equation}
where $\Lambda_g=4(1-g^2)\ll1$ near the critical point, and $\mathrm{Var}(X^{2})_{\vert\varphi\rangle}=\langle\varphi\vert X^{4}\vert\varphi\rangle-\langle\varphi\vert X^2\vert\varphi\rangle^{2}$. We stress that such a scaling of $I_{g}(t)\propto\Lambda_g^{-3}$ holds as long as $\mathrm{Var}[X^2]_{|\varphi\rangle}$ is non-vanishing, which is valid for general initial states. We assume the bosonic field is initialized in a coherent state, i.e., $|\varphi\rangle=|\alpha\rangle$, where $\alpha$ is the coherent amplitude. The dynamics of the quadrature $P$ is given by\\
\begin{eqnarray}
	\langle P\rangle_t=-\sqrt{2\eta}\alpha\sin{(\sqrt{\Lambda_g}\omega t/2)} ,
\end{eqnarray}
yielding the susceptibility with respect to the parameter $g$ as
\begin{eqnarray}
	\chi_g(\tau_n)&=&\partial_g\langle P(t)\rangle|_{t=\tau_n}=(-1)^{n-1}\frac{\sqrt{2}\alpha}{1-g}\omega\tau_ng,
\end{eqnarray}
where $\tau_n=2n\pi/(\sqrt{\Lambda_{g}}\omega)$ with $n\in\mathbb{Z}$.  A similar analysis shows that the linear dynamic range at a fixed working point narrows as $g$ approaches the critical point. This is a general feature of sensing schemes that utilize criticality and its associated divergent behavior. Indeed, the key advantage of criticality-based quantum sensing is its ability to detect minuscule changes in physical parameters. This approach is expected to have powerful applications in weak-signal detection and precision metrology. By first obtaining a relatively-accurate pre-estimation of the parameter, one can then use a bias field to tune the system close to the critical point, enhancing the sensitivity to minute variations. To determine the measurement precision, we also calculate the variance of the quadrature $P$. The variance is given by\\
\begin{align}
	(\Delta P)^{2}\!\!&=\!\!\langle P^{2}\rangle_{t}\!\!-\!\!\langle P\rangle_{t}^{2} \nonumber \\
	\!\!&=\!\!\frac{1}{2}\left[\cos^2\left({\frac{\sqrt{\Lambda_g}\omega t} {2}}\right)+\eta\sin^2\left({\frac{\sqrt{\Lambda_g}\omega t} {2}}\right)\right].
\end{align}
We note that the oscillation term in the second part of $(\Delta P)^{2}$  is out of phase with the term $ \cos(\sqrt{\Lambda_{g}}\omega t/2)$ in $\chi_g(t)$. This allows us to achieve an enhanced susceptibility while retaining a small fluctuation of the quadrature. Therefore, the measurement precision of the parameter $g$ can be significantly improved. To quantify the estimation precision, we define the inverted variance as $\mathcal{F}_{g}(t)=\chi_{g}^{2}(t)/(\Delta P)^{2}$. When $\mathcal{F}_{g}(t)\approx I_g(t)$, the precision reaches the quantum Cram$\acute{\rm e}$r-Rao bound. The local maxima of the inverted variance occur at evolution times $\tau_n$ and are given by
\begin{eqnarray}
	\mathcal{F}_{g}(\tau_n)=\frac{4n^2\pi^2\alpha^2g^2}{(1+g)(1-g)^3}.
\end{eqnarray}
Note that the inverted variance diverges in the long-time limit. The QFI at the same time is
\begin{equation}
	I_{g}(\tau_n)=16(1+g)^2\omega^2\Lambda_g^{-2}\tau_n^2\mathrm{Var}[X^2]_{|\varphi\rangle}.
\end{equation}
Notice that the local maxima $\mathcal{F}_{g}(\tau_n)$ is of the same order as the QFI $I_{g}(\tau_n)$.

\section{QUANTUM SIMULATION BY CIM}\label{sec:CIM}

In this section, we mainly describe how the CIM realizes the QPT of the LMG model and its critically-enhanced quantum sensing capability. We demonstrate that the CIM serves not only as a static simulator but also as a powerful platform for executing complex dynamical sensing protocols.

\subsection{Quantum Phase Transition in the CIM}\label{CIMtoLMGMapping}
\begin{figure}[hbtp]
	\centering
	\includegraphics[width=1\linewidth]{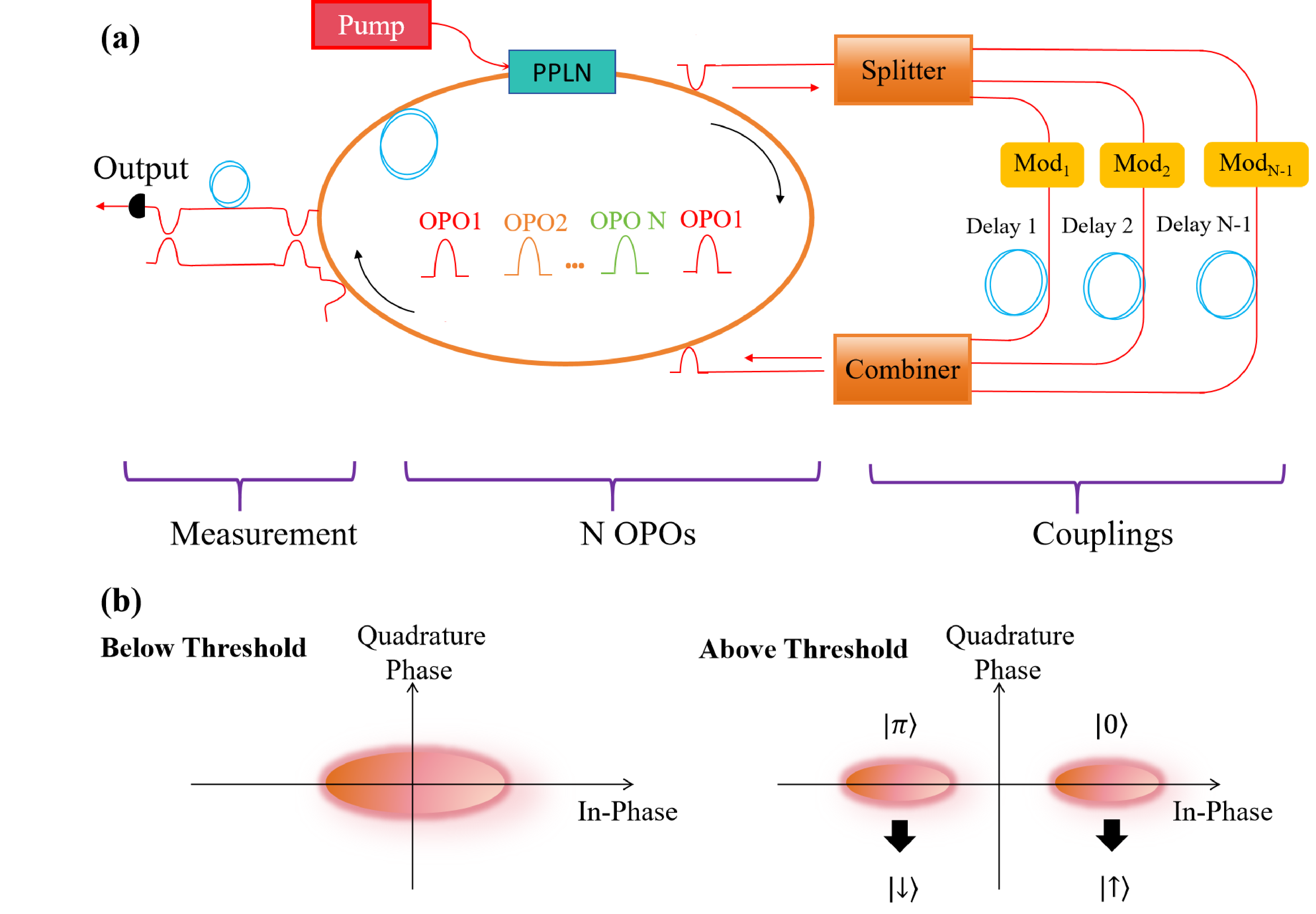} 
	\caption{Schematic of a fiber-based large-scale DOPO network. (a) A CIM based on the time-division multiplexed DOPO with mutual coupling implemented by optical delay lines. (b) Simplified illustration of vacuum-squeezing phenomenon in a DOPO below threshold and the performance of binary phases above threshold in in-phase and quadrature-phase coordinates.}
	\label{FIG:System}
\end{figure}

In our work, we employ a time-division multiplexing CIM to simulate the long-range interaction of the LMG model. The system is shown in Fig.~\ref{FIG:System}(a). Multiple DOPO pulses are generated within a single-fiber ring cavity. The signal modes of different DOPOs are mutually injected via optical delay lines, achieving all-to-all coupling among the DOPO pulses. This architecture has been successfully verified in experiments \cite{marandi2014}. Figure~\ref{FIG:System}(b) illustrates the operation of a DOPO below and above its oscillation threshold in in-phase and quadrature-phase coordinates. Below the oscillation threshold, each DOPO pulse is in a squeezed vacuum state. Above the oscillation threshold, the system evolves into a coherent state, exhibiting either a $0$-phase $|\alpha\rangle$ or $\pi$-phase state $|\!-\alpha\rangle$. We map the $0$-phase and $\pi$-phase states of each DOPO to the spin-up and spin-down states, i.e.,  $\vert\!\uparrow\rangle$ and spin-down $|\!\downarrow\rangle$, respectively. This mapping encodes the spin variables of the LMG model, enabling the DOPOs network to collectively evolve to find the state with the minimum effective Hamiltonian, thereby achieving the simulation of spin-spin interactions. The CIM can be described by the total Hamiltonian~\cite{marandi2014,haribara2016}
\begin{align}
H_{\rm tot} &= H_{\text{f}} + H_{\text{I}} + H_{\text{P}} + H_{\text{C}}. \label{eq:Hall}
\end{align}
Here,
\begin{align}
H_{\text{f}} = \sum_{j=1}^{N} (\omega_{\text{s}} a_{\text{s}j}^{\dagger}a_{\text{s}j} + \omega_{\text{p}}a_{\text{p}j}^{\dagger}a_{\text{p}j} 
+ \omega_{\text{c}}\sum_{k \neq j} a_{\text{c}jk}^{\dagger}a_{\text{c}jk})
\end{align}
is the Hamiltonian for the $j$th signal field with frequency $\omega_s$ and annihilation operator $a_{sj}$, and the $j$th pump field with frequency $\omega_p$ and annihilation operator $a_{pj}$, and the coupling field with frequency $\omega_c$ and annihilation operator $a_{cjk}$, which simultaneously couples the $j$th and $k$th signal fields.
\begin{align}
H_{\text{I}} &= \frac{i\kappa}{2} \sum_{j=1}^{N} \left( a_{\text{s}j}^{\dagger 2}a_{\text{p}j} - a_{\text{p}j}^{\dagger}a_{\text{s}j}^{2} \right) 
\label{eq42}
\end{align}
is the interaction Hamiltonian between the $j$th signal and the $j$th pump field with interaction strength $\kappa$.
\begin{align}
H_{\text{P}} &= i \varepsilon\sum_{j=1}^{N} \left(  a_{\text{p}j}^{\dagger}- a_{\text{p}j} \right) 
\end{align} 
is the external pumping Hamiltonian with real pump amplitude $\epsilon$.
\begin{align}
H_{\text{C}} &= i\zeta_{jk} \sum_{j=1}^{N} \sum_{k \neq j} \left( a_{\text{c}jk}a_{\text{s}j}^{\dagger} - a_{\text{c}jk}^{\dagger}a_{\text{s}j} \right. \notag \\ 
&\quad \left.  + a_{\text{s}k}a_{\text{c}jk}^{\dagger}e^{-i\vec{k} \cdot \vec{z}}- a_{\text{s}k}^{\dagger}a_{\text{c}jk}e^{i\vec{k} \cdot \vec{z}} \right) 
\label{eq:H_c}
\end{align}
is the coupling Hamiltonian between the signal and coupling fields with coupling strength $\zeta$. And 
\begin{align}
H_{\text{SR}} =& \sum_{j=1}^{N} \left( a_{\text{s}j}\Gamma_{\text{s}j}^{\dagger} + \Gamma_{\text{s}j}a_{\text{s}j}^{\dagger} + a_{\text{p}j}\Gamma_{\text{p}j}^{\dagger} + \Gamma_{\text{p}j}a_{\text{p}j}^{\dagger} \right) \notag  \nonumber \\ 
& +  \sum_{j=1}^{N} \sum_{k \neq j} \left( a_{\text{c}jk}\Gamma_{\text{c}}^{\dagger} + a_{\text{c}jk}^{\dagger}\Gamma_{\text{c}} \right)
\end{align}
is the system-reservoir interaction Hamiltonian which describes any dissipation processes for the signal, pump and coupling fields. In the coupling Hamiltonian $H_{\rm C}$, 
The phase factors $\exp(\pm i\vec{k} \cdot \vec{z})$ represent the in-phase or out-of-phase coupling from the $j$-th DOPO to the $k$-th DOPO pulse. Different couplings can be obtained based on the value of $\exp(ik_{c}z)$ and $\exp(-ik_{c}z)$. The two dominant terms in the Hamiltonian $H_{\rm tot}$ are the parametric coupling term $H_{\rm I}$ and the mutual coupling term $H_{\rm C}$ between different pulses. The former generates squeezing in each DOPO pulse, while the latter introduces tunable interactions between pulses, effectively mapping the LMG Hamiltonian onto the network dynamics. The corresponding Heisenberg-Langevin equations for a single DOPO are
\begin{eqnarray}
\frac{da_p}{dt} &=& - \gamma_p a_p+\varepsilon - \frac{\kappa}{2} a_s^2+\sqrt{\gamma_p}\xi_1, \label{eq7.1}\\
\frac{da_s}{dt} &=& - \gamma_s a_s + \kappa a_s^{\dagger} \hat{a}_p+\sqrt{\gamma_s}\xi_2,
	\label{eq7}
\end{eqnarray}
where $\xi_i$ ($i=1,2$) is a quantum-noise operator satisfying $\langle{\xi_i^\dagger(t)}{\xi_i^\dagger(t')}\rangle=0$ and  $\langle{\xi_i(t)}{\xi_i^\dagger(t')}\rangle=2\delta_{ij}\delta(t-t')$. The parameters $\gamma_p$ and $\gamma_s$ represent the relaxation rates for the pump and signal light, respectively.
The oscillation threshold is $\epsilon_{\rm th}=\gamma_p\gamma_s/\kappa$. If $\gamma_p\gg\gamma_s$, we can adiabatically eliminate the pump mode by setting $da_p/dt=0$, yielding
\begin{equation}
a_p = \frac{\varepsilon}{\gamma_p} - \frac{\kappa}{2 \gamma_p} a^2_s + \sqrt{\frac{1}{\gamma_p}}\xi_1.
\label{eq9}
\end{equation}
{We note that Eq.~\eqref{eq9} contains a Langevin noise contribution from the pump mode. In the derivation of the effective Hamiltonian, we ignore this noise term to obtain a compact analytic mapping. The validity of this adiabatic elimination and the role of the pump noise are analyzed in Appendix~\ref{appendixA}, where we keep the noise term explicitly and analyse the simplified description against the full stochastic dynamics as seen in Fig.~\ref{FIG:SM}.} Here, by ignoring the noise term in the above formula and then substituting it into Eq.~(\ref{eq42}), we can obtain
\begin{eqnarray}
H_{\rm I}=i\frac{\kappa\varepsilon}{2\gamma_p}\sum_{j=1}^{N}(a_{sj}^{\dagger2}-a_{sj}^{2})=i\sum_{j=1}^{N} \frac{S}{2} ( a_{j}^{\dagger 2} - a_{j}^{2} ),
\end{eqnarray}
where $S=\kappa\epsilon/\gamma_p$ is a squeezing parameter, and we have removed the index s of the signal mode in the subscript for convenience. Similarly, adiabatically eliminating the coupling field $a_{cjk}$ by setting $da_{cjk}/dt=0$ and selecting a phase $k_{c}z=\pi/2$ results in
\begin{eqnarray}
 a_{cjk} = -\frac{\zeta_{jk} }{\gamma_{c}}\left( a_{sj} + ia_{sk} \right). \label{eq:a_{cjk}}
\end{eqnarray}
{Similarly, the elimination of the coupling fields $a_{cjk}$ is validated by a separation of time scales, i.e., $\gamma_c \gg \gamma_s$, as discussed in Appendix~\ref{appendixA}.} Substituting it into Eq.~(\ref{eq:H_c}) yields an effective fully-connected dual-mode squeezing interaction
\begin{eqnarray}
H_{\text{C}}^{\rm eff} = - \frac{J}{N} \sum_{1\leq j<k\leq N}^{N} \left( a_{sj} a_{sk} +  a_{sk}^{\dagger}a_{sj}^{\dagger} \right),\label{eq:HC}
\end{eqnarray}
where $J/N=J_{jk}=\zeta^2_{jk}/\gamma_c$ is chosen to ensure uniform coupling strength across all pairs. Thus, we can obtain the effective total Hamiltonian
\begin{align}
H_{\text{eff}}=& \Delta\sum_{j=1}^{N}a_{j}^{\dagger}a_{j}+i\sum_{j=1}^{N} \frac{S}{2} \left( a_{j}^{\dagger 2} - a_{j}^{2} \right) \nonumber \\
&- \frac{J}{N}\sum_{1\leq j<k\leq N}^{N} \left(a_{j}a_{k} + a_{k}^{\dagger}a_{j}^{\dagger} \right)\label{eq:Heff},
\end{align}
where $\Delta=\omega_s-\omega_p/2$ is detuning. To simulate the LMG model, it requires not only the construction of its specific spin-spin interactions but also the generation of a transverse field, which in this scheme can be effectively achieved using free-energy terms and local squeezing terms. To this end, we apply a unitary transformation
\begin{align}
U = \prod_{j=1}^{N} \exp\left( \frac{1}{2}\xi^{*}a_{j}^{2} - \frac{1}{2}\xi a_{j}^{\dagger 2} \right),
\end{align}
with $\xi=r\exp(i\theta)$ to the system. We choose the phase $\theta=\pi/2$ to set $\xi=ir$. Thus, we have
\begin{align}
U = \exp\left[ -i\frac{r}{2}\sum_{j=1}^{N} \left( a_{j}^{2} + a_{j}^{\dagger 2} \right) \right].
\end{align}
This transformation yields
\begin{align}
a_{j} &= b_{j}\cosh r - i b_{j}^{\dagger}\sinh r. 
\end{align}
Substituting it into Eq.~(\ref{eq:Heff}) and choosing the parameter $r$ to ensure that the coefficients of the $a^2$ and $a^{\dagger 2}$ vanish, we obtain
\begin{align}
H_{\text{eff}}' = \Omega \sum_j b_j^{\dagger} b_j - \frac{J}{N} \sum_{j<k} \left( b_j b_k + b_k^{\dagger}b_j^{\dagger} \right),\label{eq:H57}
\end{align}
where $\Omega=\sqrt{\Delta^2-S^2}$ is the renormalized oscillation frequency. 
Under the low-excitation approximation, 
we now introduce the pseudospin operators via the mapping \cite{Matsubara1956PTP}
\begin{subequations}\label{eq:47}
\begin{align}
b_j &\to \sigma_j^- = \frac{1}{2}(\sigma_j^x - i\sigma_j^y), \label{eq:47a} \\ 
b_j^\dagger &\to \sigma_j^+ = \frac{1}{2}(\sigma_j^x + i\sigma_j^y), \label{eq:47b} \\ 
b_j^\dagger \hat{b}_j &\to \frac{1}{2}(1 + \sigma_j^z). \label{eq:47c}
\end{align}
\end{subequations}
Substituting these into Eq.~(\ref{eq:H57}) and neglecting constant terms, we can obtain the final form of the generalized LMG model Hamiltonian
\begin{align}
H_{\text{fin}}=-\frac{J}{2N}\sum_{1\leq j<k\leq N}^{N} \left( \sigma_{j}^{x}\sigma_{k}^{x} - \sigma_{j}^{y}\sigma_{k}^{y} \right) + \frac{\Omega}{2}\sum_{j=1}^{N} \sigma_{j}^{z},
\end{align}
where the parameters of the simulated LMG model are related to the CIM parameters by $\lambda=-J/2$ and $h=\Omega/2$. Thus, the CIM can serve as a quantum simulator of the LMG model, where one can simulate a wide range of coupling regimes by suitably tuning the laser intensities and detunings to match the desired ratio. This is also a unique advantage of the CIM platform over other quantum simulators when studying such long-range interaction models. The corresponding parameter relationships are shown in Tab.~\ref{tab:CIM and LMG}.
\begin{table}[h]
	\centering
	\caption{Parameter correspondence}
	\fontsize{8pt}{16pt}\selectfont
	\setlength{\tabcolsep}{18pt}
	\renewcommand{\arraystretch}{2}
	\begin{tabular}{|c|c|} 
		\hline
		CIM & LMG \\  
		\hline
		\(\displaystyle \Omega=\sqrt{\Delta^2-S^2}=\sqrt{\Delta^2-\left(\frac{\kappa\epsilon}{\gamma_p}\right)^2}\) & \(\displaystyle h=\frac{\Omega}{2}\) \\  
		\hline
		\(\displaystyle \frac{J}{N}=\frac{\zeta^2_{jk}}{\gamma_c}\) & \(\displaystyle \lambda=-\frac{J}{2}\) \\
		\hline
	\end{tabular}
	\label{tab:CIM and LMG}
\end{table}

Introducing the collective spin operators $S_{\alpha}=\sum_{j=1}^{N}\sigma_{j}^{\alpha}/2$ ($\alpha=x,y,z$), we can obtain
\begin{equation}
H_{\text{fin}}=-\frac{J}{N}\left(S_{x}^{2}-S_{y}^{2}\right)+\Omega S_{z}.
\label{eq:H62}
\end{equation}
Here, $J<0$ corresponds to the anti-ferromagnetic coupling and $J>0$ to the ferromagnetic coupling. This demonstrates that the CIM can simulate the LMG model for both ferromagnetic and anti-ferromagnetic coupling. Using the Holstein-Primakoff and Bogoliubov transformation as in Sec.~\ref{sec:LMG}, the Hamiltonian is diagonalized as
\begin{align}
	H_{\rm CIM} = 
	\begin{cases} 
		E_{k1}b^\dagger b+E_{g1}, & |\Omega| \geq J, \\
		E_{k2}b^\dagger b+E_{g2}, & |\Omega| < J,
	\end{cases}
\end{align}
with the eigen energies and ground-state energies being respectively
\begin{subequations}\label{eq:51}
\begin{align}
E_{k1}&=\sqrt{\Omega^2-J^2}, \\
E_{g1}&=\frac{1}{2}\sqrt{\Omega^2-J^2}-\frac{1}{2}\Omega(N+1), \label{eq:51a} \\
E_{k2}&= \sqrt{2}\sqrt{J^2-\Omega^2}, \label{eq:51b} \\
E_{g2}&=\frac{\sqrt{2}}{2}\sqrt{J^2-\Omega^2}-\frac{3J}{4}+\frac{\Omega^2}{4J}-\frac{JN}{4}-\frac{\Omega^2N}{4J}. \label{eq:51c}
\end{align}
\end{subequations}
Thus, the rescaled ground-state energy is
\begin{align}
e_g \equiv \frac{E_g}{N}= 
\begin{cases} 
-\frac{\Omega}{2}, & |\Omega| \geq J, \\
-\frac{J}{4} - \frac{\Omega^2}{4J}, & |\Omega|<J.
\end{cases}
\label{eq:CIMeg}
\end{align}
This clearly shows the non-analyticity at the quantum critical point $|\Omega|=J$, or equivalently $h=\pm\lambda$, confirming that the CIM can faithfully reproduce the second-order QPT of the LMG model. The proposed simulation scheme is well within the reach of the current CIM architectures. For instance, the time-multiplexed CIM with all-to-all couplings have already been demonstrated with up to 100,000 nodes \cite{SA100000}, far exceeding the system sizes required for observing the thermodynamic-limit behavior of the LMG model. The tunable coupling strengths $J_{jk}$ can be implemented via adjusting optical injection phases and amplitudes, allowing precise control over the simulated Hamiltonian parameters.

\subsection{Quantum Metrology of CIM}

In our scheme, the signal DOPO mode encodes the spin configurations, consistent with standard CIM operation. To study the ability of CIM to simulate the criticality-enhanced quantum sensing of the LMG model, following Eqs.~(\ref{eq:3a}-\ref{eq:4c}), we can obtain the relevant Hamiltonian
\begin{align}
H=\Omega\left[a^\dagger a-\frac{\tilde{g}}{2}(a^2+a^{\dagger 2})\right],
\label{eq:DOPO}
\end{align}
where $\tilde{g}=J/\Omega=-g$. Using Eq.~(\ref{eq:XY}), we can obtain
\begin{align}
H=\frac{\Omega(1+\tilde{g})}{2}\left[P^2+\left(1-\frac{2\tilde{g}}{1+\tilde{g}}\right)X^2\right].
\end{align}
Let $H_0=\Omega(1+\tilde{g}) P^2/2$ and $H_1=\Omega(1+\tilde{g}) X^2/2$. The QFI for estimating the parameter $\tilde{g}$ near the critical point can be expressed as
\begin{equation}
I_{\tilde{g}}(t)\simeq16(1-\tilde{g})^2\frac{[\sin(\sqrt{\Lambda_{\tilde{g}}}\Omega t)-\sqrt{\Lambda_{\tilde{g}}}\Omega t]^{2}}{\Lambda_{\tilde{g}}^{3}}\mathrm{Var}[X^2]_{|\varphi\rangle},\label{eq:CIMI}
\end{equation}
where $\Lambda_{\tilde{g}}=4(1-\tilde{g}^2)$, and $|\varphi\rangle=|\alpha\rangle$ is the same as the LMG model. In order to encode spin variables in DOPO pulses, the CIM should operate above the threshold. Thus, it is reasonable to set $|\varphi\rangle$ as the coherent state $|\alpha\rangle$. After evolving under the Hamiltonian~(\ref{eq:DOPO}) for an interal $t$, we perform measurements for the quadrature $P$. Its mean and variance are respectively
\begin{align}
\langle P\rangle_t
\!&=\!-\sqrt{2\eta^{\prime}}\alpha\sin{(\sqrt{\Lambda_{\tilde{g}}}\Omega t/2)}, \\
(\Delta P)^{2}\!&=\!\frac{1}{2}\left[\cos^2{(\sqrt{\Lambda_{\tilde{g}}}\Omega t/2)}\!+\!\eta^{\prime}\sin^2{(\sqrt{\Lambda_{\tilde{g}}}\Omega t/2)}\right],
\end{align}
where $\eta^{\prime}=1-2\tilde{g}/(1+\tilde{g})$.
With the quadrature $P$ serving as the sensing indicator, the susceptibility, i.e., $\chi_{\tilde{g}}(t)=\partial_{\tilde{g}}\langle P(t)\rangle$, becomes divergent in the vicinity of the critical point. To quantify the estimation precision, we define the inverted variance $\mathcal{F}_{\tilde{g}}(t)=\chi_{\tilde{g}}^{2}(t)/(\Delta P)^{2}$, whose upper bound is imposed by the quantum Cram\'{e}r-Rao bound, i.e., $\mathcal{F}_{\tilde{g}}(t)\leq I_{\tilde{g}}(t)$. The inverted variance $\mathcal{F}_{\tilde{g}}(t)$ is a periodic function of the evolution times $\tau_n=2n\pi/(\sqrt{\Lambda_{\tilde{g}}}\Omega) (n\in\mathbb{Z})$ with the local maxima
\begin{eqnarray}
\mathcal{F}_{\tilde{g}}(\tau_n)=\frac{4n^2\pi^2\alpha^2{\tilde{g}}^2}{(1-{\tilde{g}})(1+\tilde{g})^3}.\label{eq:CIMF}
\end{eqnarray}
The QFI at the time $\tau_n$ is
\begin{equation}
I_{\tilde{g}}(\tau_n)=16(1+\tilde{g})^2\Omega^2\Lambda_{\tilde{g}}^{-2}\tau_n^2\mathrm{Var}[X^2]_{|\varphi\rangle}.
\end{equation}
The divergent behavior of both $\mathcal{F}_{\tilde{g}}$ and $I_{\tilde{g}}$ as $\tilde{g}\rightarrow 1$ demonstrates that the criticality-enhanced measurement precision is achievable in the CIM platform when simulating the LMG model near its quantum critical point.

\section{Results and Discussion}\label{sec:Results}

In Sec.~\ref{sec:LMG} and ~\ref{sec:CIM}, we have established the theoretical framework for the QPT and critically-enhanced quantum sensing in the LMG model, and further proposed a scheme to realize the LMG Hamiltonian in a CIM. In this section, we present numerical and analytical results to demonstrate the QPT and quantum sensing performance of the LMG model, and validate the capability of the CIM to emulate both the QPT and the quantum sensing behavior of the LMG model.

\subsection{Result analysis of quantum phase transition}\label{sec:Result of the QPT}

Our exact solutions show that the QPT of the LMG model occurs at the critical point $h_c=1$. The rescaled ground-state energy, i.e., $e_g\equiv E_g/N$, is $-\lambda/2-h^2/(2\lambda)$ for $h<h_c$ and $-h$ for $h\geq h_c$. For the CIM, the QPT occurs at $\Omega_c=\pm J$, and its rescaled ground-state energy is $e_g=-J/4-\Omega^2/(4J)$ for $|\Omega|<J$ and $e_g=-\Omega/2$ for $|\Omega|\geq J$. 
\begin{figure}[hbtp]
\centering
\includegraphics[width=1\linewidth]{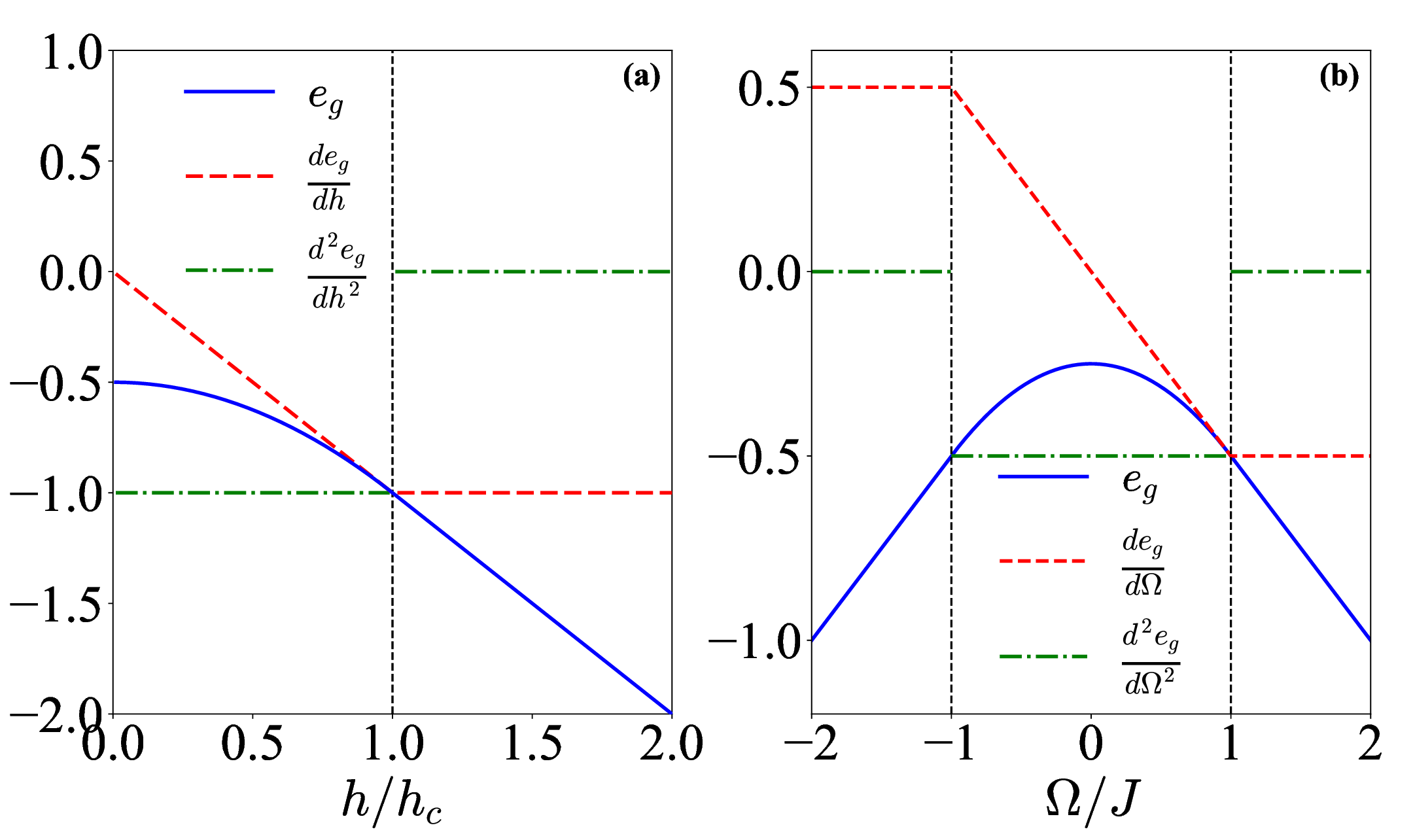}
\caption{Comparison of exact solutions for the LMG model and CIM. (a) Exact solutions of the LMG model as a function of the magnetic-field strength $h/h_c$ for the rescaled ground-state energy $e_g$ (blue solid line), $de_g/dh$ (red dashed line) and $d^2e_g/dh^2$ (green dashed-dotted line). (b) Exact solutions of the CIM as a function of the renormalized oscillation frequency $\Omega/J$ for the rescaled ground-state energy $e_g$ (blue solid line), $de_g/d\Omega$ (red dahed line) and $d^2e_g/d\Omega^2$ (green dashed-dotted line).}
\label{FIG:QPT}
\end{figure}
Figure~\ref{FIG:QPT}(a) shows the exact solutions of the LMG model as a function of the magnetic-field strength $h/h_c$ for the rescaled ground-state energy $e_g$, its first derivative $de_g/dh$, and second derivative $d^2e_g/dh^2$. One can see that $e_g$ and $de_g/dh$ are continuous, while $d^2e_g/dh^2$ exhibits an abrupt jump at $h=h_c$. This non-analytic behavior reveals the second-order nature of the QPT. In contrast, Fig.~\ref{FIG:QPT}(b) shows the corresponding exact solutions for the CIM as a function of $\Omega/J$ for the rescaled ground-state energy $e_g$, $de_g/d\Omega$ and $d^2e_g/d\Omega^2$. The behavior of the CIM at the critical point $\Omega/J=1$ is highly consistent with that of the LMG model for $\Omega/J>0$, indicating that the mapping of the LMG spins to DOPO phases and the encoding of long-range interactions by optical mutual injection allow the CIM to effectively reproduce the critical phenomena of the LMG model. Furthermore, the QPT exhibited by the CIM at $\Omega/J=-1$ accurately corresponds to the critical behavior of the LMG model at $h/\lambda=-1$. This result reflects the intrinsic $\mathbb{Z}_2$ symmetry of the LMG Hamiltonian, i.e., its invariance under a simultaneous $\pi$ rotation of all spins around the $z$-axis. This symmetry implies two equivalent critical points at $h=\pm \lambda$, marking the transition from the paramagnetic phase to two possible symmetry-broken phases. 

The successful simulation by the CIM theoretically validates this symmetry, i.e., as $\Omega/J\rightarrow-1$, the derivatives of the ground-state energy exhibit singular behavior identical to that for $\Omega/J\rightarrow1$, strongly supporting the physical equivalence of these two critical points. Therefore, the CIM does not only reproduce the critical behavior at a single point, but can also draw the entire phase diagram of the LMG model by changing its parameter range, revealing its potential symmetrical structure. {Furthermore, the dissipation enters the signal modes at the Langevin level and modifies the effective control parameters through renormalization.
In our mapping, the renormalized oscillation frequency is $\Omega=\sqrt{\Delta^2-S^2}$ with $S=\kappa\epsilon/\gamma_p$. Thus, the dissipation-dependent operating conditions, e.g., above-threshold pumping with $\epsilon\propto \gamma_s$, shift the apparent critical detuning. Increasing $\gamma_s$ and hence $S$ primarily shifts the apparent critical point in detuning, i.e., $\Delta_c=\sqrt{J^2+S^2}$, and thus changes the magnitude of the derivatives with respect to $\Delta$. In practice, any ideal divergences predicted in the thermodynamic and long-time limits are regularized by finite interrogation time and finite system size $N$.
}

\subsection{Result analysis of quantum metrology}

\begin{figure*}[hbtp]
\centering
\includegraphics[width=1\linewidth]{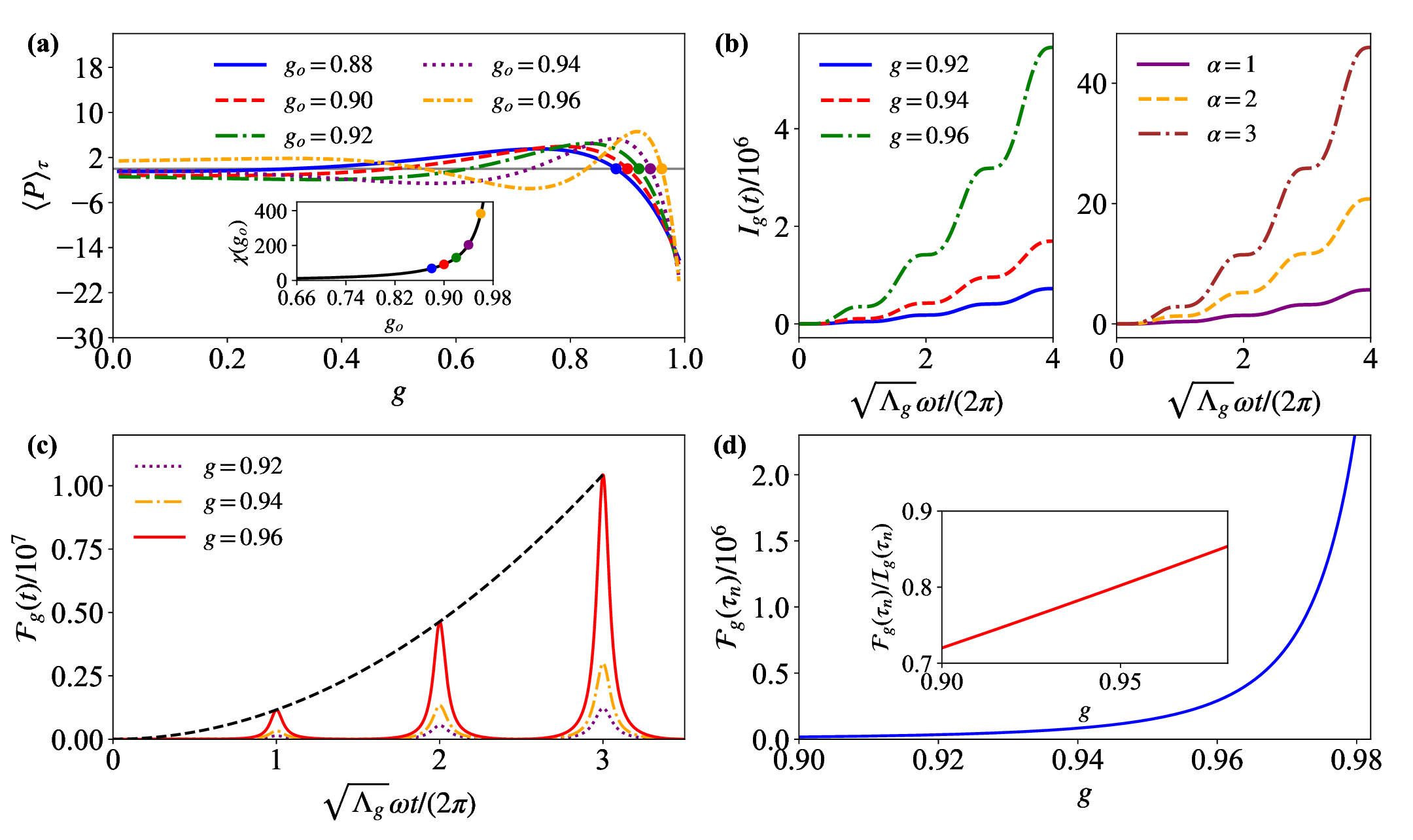}
\caption{The numerical evaluation results of the QPT and criticality-enhanced quantum sensing of the LMG model. (a) Quadrature $\langle\hat{P}\rangle_\tau$ of the LMG model after an evolution time $\tau_n=\pi/[\omega(1-g^2_o)^{1/2}]$ as a function of $g$ with $\alpha=1$. The working point $g=g_o$ is marked by filled circles. The inset shows the corresponding susceptibility at the working point $g_o$. (b) The variation relationship of the QFI in the LMG model under different parameters $g$ and coherent amplitude $\alpha$. (c) The inverted variance $\mathcal{F}_g(t)$ of the LMG model vs the evolution time $t$. $\mathcal{F}_g(t)$ shows equidistant peaks at $\sqrt{\Lambda_g}\omega t/(2\pi)$. (d) The peak of the inverted variance $\mathcal{F}_g(t)$ of the LMG model vs the parameter $g$ for an evolution time $\tau_n$ with $\alpha=1$. The inset shows that the local maxima of $\mathcal{F}_g(\tau_n)$ reaches  the seam order of $I_g(\tau_n)$.}
\label{FIG:LMG}
\end{figure*}

In the Sec.~\ref{sec:Result of the QPT}, we analyzed the QPT behavior of the LMG model and CIM at the critical point. Now, we are very interested in studying their quantum metrology at the critical point. We begin by calculating the dynamics of the quadrature observable $\langle P\rangle_\tau$ in the LMG model as a function of the parameter $g$. One can see from the Fig.~\ref{FIG:LMG}(a) that $\langle P\rangle_\tau$ becomes very sensitive to the change of $g$ near the critical point $g_c=1$. The result shows that as the working point $g_o$ approaches the critical point, the slope of $\langle P\rangle_\tau$ at $g_o$ diverges significantly, while the linear dynamic range around $g_o$ narrows. This indicates that the system achieves higher sensitivity to a small change in $g$ around the criticality, albeit within a narrower parameter range.

The inset of Fig.~\ref{FIG:LMG} (a) plots the susceptibility $\chi(g_o)$ as a function of the working point $g_o$. The susceptibility $\chi(g_o)$ exhibits divergent behaviors as $g_o\rightarrow1$, i.e., $\Lambda_g\rightarrow0$, demonstrating the critical enhancement of the response. The balance between the enhanced sensitivity and the narrowed dynamic range is a common characteristic of the criticality-enhanced sensing schemes. This requires relatively-accurate estimation of parameters, so that the system is biased near the critical point. After that, its enhanced sensitivity can be used for the weak-signal detection or the precise measurement. The quantum limit of this precision is decided by the QFI. The first subimage of the Fig.~\ref{FIG:LMG} (b) plots the QFI $I_g(t)$ of the LMG model as the function of the evolution time $t$ for different values of the parameter $g$ near the critical point with the coherent amplitude $\alpha=1$. The result indicates that the QFI $I_g(t)$ exhibits oscillatory behavior. Crucially, its amplitude significantly increases as $g\rightarrow1$, which is a direct result of the closure of the energy gap, i.e., $\Lambda_g\rightarrow0$, slowing down the system's evolution and amplifying the quantum state's response to parameter changes. This trend reveals that the system is very sensitive to small changes in parameter $g$ near the critical point, and the periodic oscillations in dynamic evolution provide the possibility for optimizing the measurement-time window. Furthermore, as shown in second subimage of the Fig.~\ref{FIG:LMG} (b), the QFI is enlarged as the increase of the initial coherent amplitude $\alpha$, indicating that the QFI can be amplified by using a coherent state with a larger amplitude $\alpha$ as the probe, and confirming that the initial coherent amplitude is a valuable resource for quantum sensing.

The achievable precision in a specific measurement protocol is given by the inverted variance $\mathcal{F}_g(t)$ of the quadrature $P$. Figure~\ref{FIG:LMG} (c) shows $\mathcal{F}_g(t)$ of the LMG model as a function of the time $t$ for different values of the parameter $g$. One can see that the $\mathcal{F}_g(t)$ gradually rises with the increase of evolution time $t$, and reaches periodic maxima at the optimal measurement time $\tau_n$. We can also see that $\mathcal{F}_g(t)$ exhibits multiple peaks as the evolution time $t$ increases, and the longer the evolution time $t$ is, the larger the peaks appear. The dashed line in Fig.~\ref{FIG:LMG} (c) represents the fitting results of these peaks over time for $g = 0.96$, yielding $\mathcal{F}_g(t) \propto t^2$, which indicates that the criticality-enhanced protocol has reached the Heisenberg limit. Similar scaling behavior is observed for other values of $g$ near the critical point. Although implementing this protocol requires selecting an evolution time near one of these peaks, Fig.~\ref{FIG:LMG} (c) shows that the peaks possess a finite width, thereby offering a degree of robustness against small timing errors.

Figure~\ref{FIG:LMG} (d) plots the peak of the inverted variance $\mathcal{F}_g(\tau_n)$ of the LMG model as a function of the parameter $g$. It can be observed that as the parameter $g$ approaches the critical point  $g_c$, the inverse variance $\mathcal{F}_g(\tau_n)$ exhibits a sharp divergent behavior, demonstrating the critical enhancement of measurement precision. The inset demonstrates that $\mathcal{F}_g(\tau_n)$,  reaches the same order of magnitude as the QFI $I_g(\tau_n)$, proving that the simple homodyne measurement of the quadrature $P$ is an effective pointer for evaluating the control parameter $g$.
\begin{figure}[hbtp]
\centering
\includegraphics[width=1\linewidth]{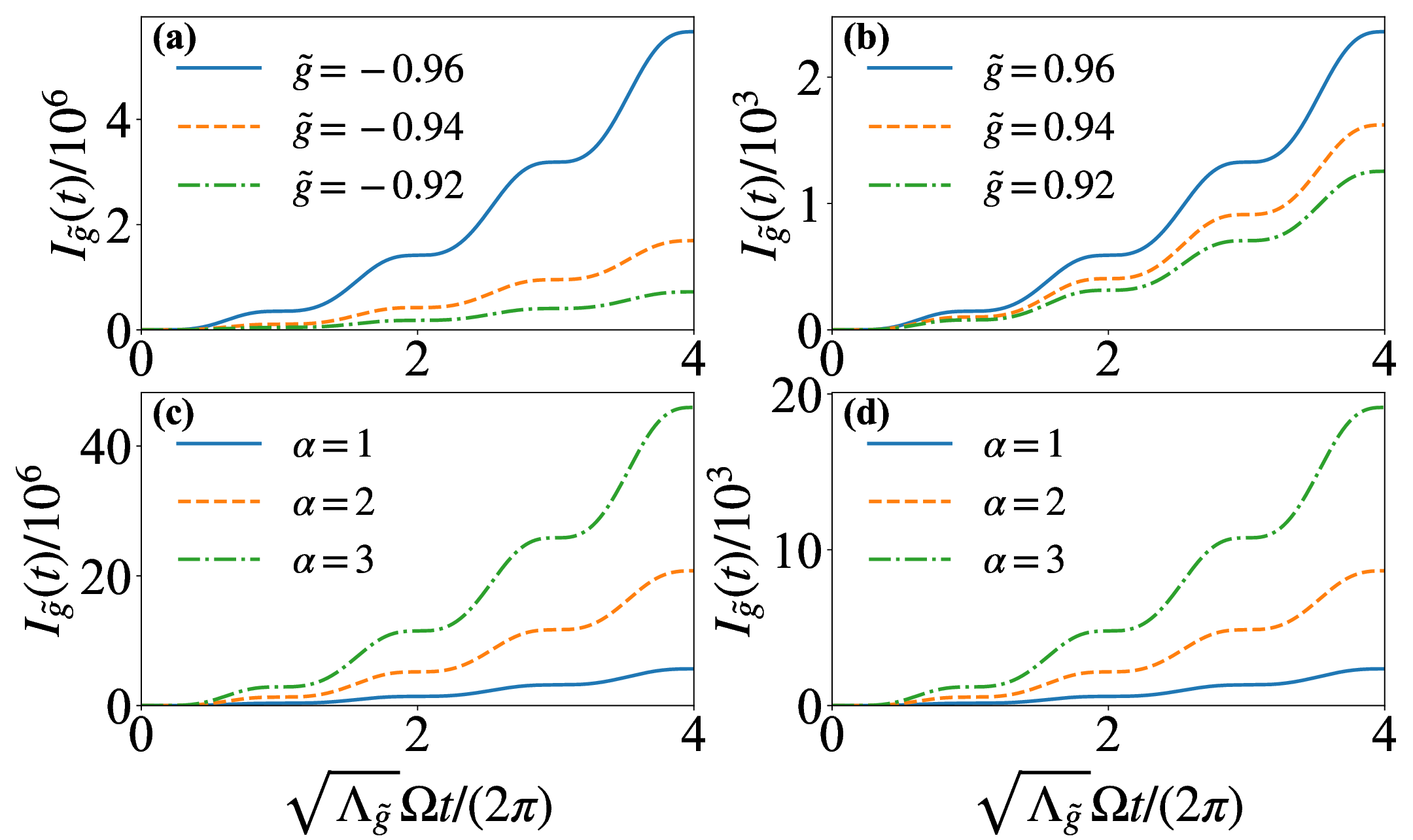}
\caption{The QFI $I_{\tilde{g}}(t)$ of the CIM as a function of the evolution time $t$ for (a) $\tilde{g}=-0.96, -0.94, -0.92$ with $\alpha=1$, (b) $\tilde{g}=0.96, 0.94, 0.92$ with $\alpha=1$, (c) $\alpha=1, 2, 3$ with $\tilde{g}=-0.96$, and (d) $\alpha=1, 2, 3$ with $\tilde{g}=0.96$.}
\label{FIG:CIMIg}
\end{figure}

{After establishing the criticality-enhanced quantum metrology in the LMG model, we now resort to these results in the CIM by using the parameter mapping derived in Sec.~\ref{CIMtoLMGMapping}. 
Specifically, starting from the CIM dynamics, we perform adiabatic elimination and the pseudo-spin mapping to obtain an effective Hamiltonian that is close to the LMG form, with the effective couplings expressed explicitly in terms of the CIM parameters, e.g., the pump strength $\epsilon$, the detuning $\Delta$, the loss rates $\gamma_{p/c}$, and the mutual-injection coupling $J$.
Therefore, the CIM curves reported below are the CIM-parameterized predictions of the mapped effective Hamiltonian, i.e., the same analytical procedure as for the LMG model, but evaluated at parameters determined by the CIM. Figure~\ref{FIG:CIMIg} shows the time evolution of the QFI $I_{\tilde{g}}(t)$ evaluated for the CIM-parameterized effective Hamiltonian, for several values of the equivalent coupling $\tilde g$ and coherent amplitude $\alpha$.} As clearly shown in Fig.~\ref{FIG:CIMIg}(a) and (c), the CIM fully reproduces the temporal oscillation of the QFI and its criticality-enhanced behavior near the critical point, consistent with the results for the LMG model presented in Fig.~\ref{FIG:LMG} (b). This confirms that the CIM has successfully reproduced the key metrological feature of the LMG model, i.e., the divergent QFI at the critical point. By adjusting the pump strength and detuning, the CIM can make its equivalent parameter $\tilde{g}$ cross the critical point generating the same critical enhancement effect as the LMG model. Furthermore, Fig.~\ref{FIG:CIMIg}(b) and (d) provide corresponding results for $g<0$, i.e., the ferromagnetic coupling case, demonstrating that the LMG model still exhibits criticality-enhanced capability in quantum sensing even under ferromagnetic coupling. Although the enhancement in this case is significantly weaker compared to the antiferromagnetic coupling case, i.e., $\tilde{g}>0$, both share the same characteristic that the measurement precision can be further improved by increasing the coherent-state amplitude $\alpha$. {Meanwhile, in the CIM, the dimensionless metrological parameter is chosen as $\tilde{g} = J/\Omega$ consistent with the LMG mapping $g=\lambda/h=-J/\Omega$. Here $J$ is the effective mutual-injection coupling, while $\Omega=\sqrt{\Delta^2-S^2}$ is the renormalized oscillation frequency determined by the cavity detuning $\Delta$ and the pump-dependent squeezing parameter $S=\kappa\epsilon/\gamma_p$. Therefore, estimating $\tilde{g}$  can be given a direct physical interpretation. For instance, an external perturbation that shifts the optical detuning $\Delta$, e.g., the refractive-index, leads to a measurable change in $\Omega$, and thus in $\tilde{g}=J/\Omega$ after calibrating $J$ and $S$. Alternatively, perturbations of the mutual-injection path can be mapped onto changes in $J$, again yielding a meaningful sensing task in the underlying optical platform.}\\
\begin{figure}[hbtp]
\centering
\includegraphics[width=1\linewidth]{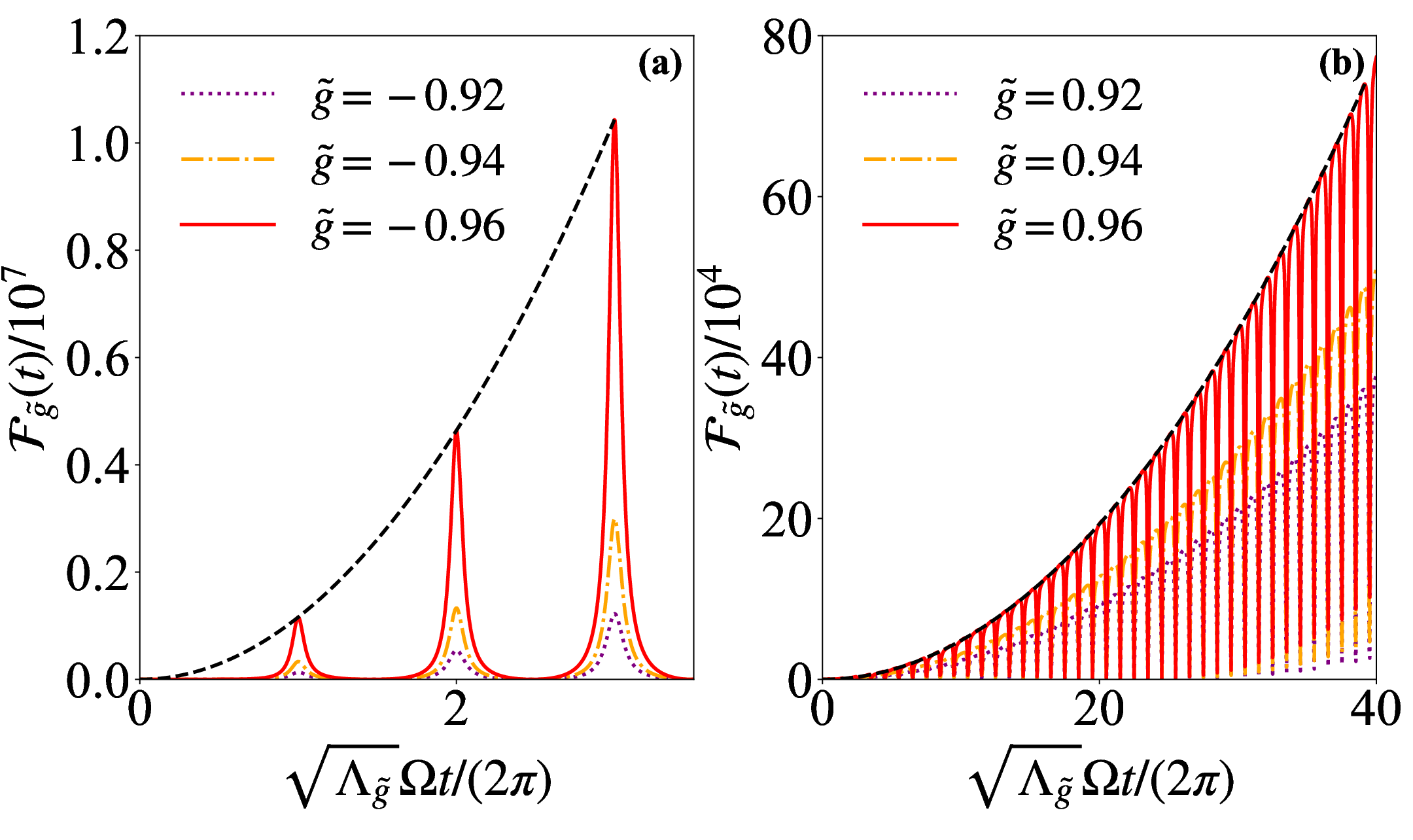}
\caption{The inverted variance $\mathcal{F}_{\tilde{g}}(t)$ of the CIM as a function of the evolution time $t$ for (a) $\tilde{g}=-0.92, -0.94, -0.96$ with $\alpha=1$ and (b) $\tilde{g}=0.92, 0.94, 0.96$ with $\alpha=1$}
\label{FIG:CIMFg}
\end{figure}
Figure~\ref{FIG:CIMFg} shows the time evolution of the inverse variance $\mathcal{F}_{\tilde{g}}(t)$ in the CIM for different parameter values of $\tilde{g}$. Figure~\ref{FIG:CIMFg}(a) shows that for the antiferromagnetic case, i.e., $\tilde{g}<0$, $\mathcal{F}_{\tilde{g}}(t)$ exhibits the same dynamic characteristics as the LMG model shown in Fig.~\ref{FIG:LMG} (c), i.e., both show a series of peaks occurring at times determined by $\sqrt{\Lambda_{\tilde{g}}}\Omega t/(2\pi)$, with amplitudes growing over time $t$. However, for the ferromagnetic case, i.e., $\tilde{g}>0$ in Fig.~\ref{FIG:CIMFg}(b), although $\mathcal{F}_{\tilde{g}}(t)$ also shows periodic peaks that increase over time, it requires a longer evolution time to achieve the same precision scaling $\mathcal{F}_{\tilde{g}}(t)\propto t^2$ as in the antiferromagnetic case. The comparison indicates that the criticality-induced enhancement of the inverted variance is substantially stronger under antiferromagnetic coupling $g<0$ than under ferromagnetic coupling $g>0$ in the LMG model, highlighting the significant influence of the coupling type on the metrological performance. {The results obtained by numerically evaluating the CIM-derived effective LMG Hamiltonian} show that under the antiferromagnetic coupling condition $\tilde{g}<0$, the divergence of the inverse variance is significantly stronger than that under the ferromagnetic coupling condition $\tilde{g}>0$. This further demonstrates that the type of coupling in the LMG model plays a crucial role in the quantum-criticality-enhancement effect. {Furthermore, the susceptibility and related figures of merit exhibit peaks at the optimal interrogation times $\tau_n$ in Eq.~\eqref{eq:CIMF}. In practice, the maximum $n$ is limited by the available interrogation time $T$ set by the cavity lifetime, the loss, and data-acquisition constraints, i.e., one must choose $\tau_n\le T$. If $T$ is shorter than the characteristic timescale $1/(\sqrt{\Lambda_g}\omega)$, the system cannot reach the peak response and the enhancement is correspondingly simplified.}

\begin{figure}[hbtp]
\centering
\includegraphics[width=1\linewidth]{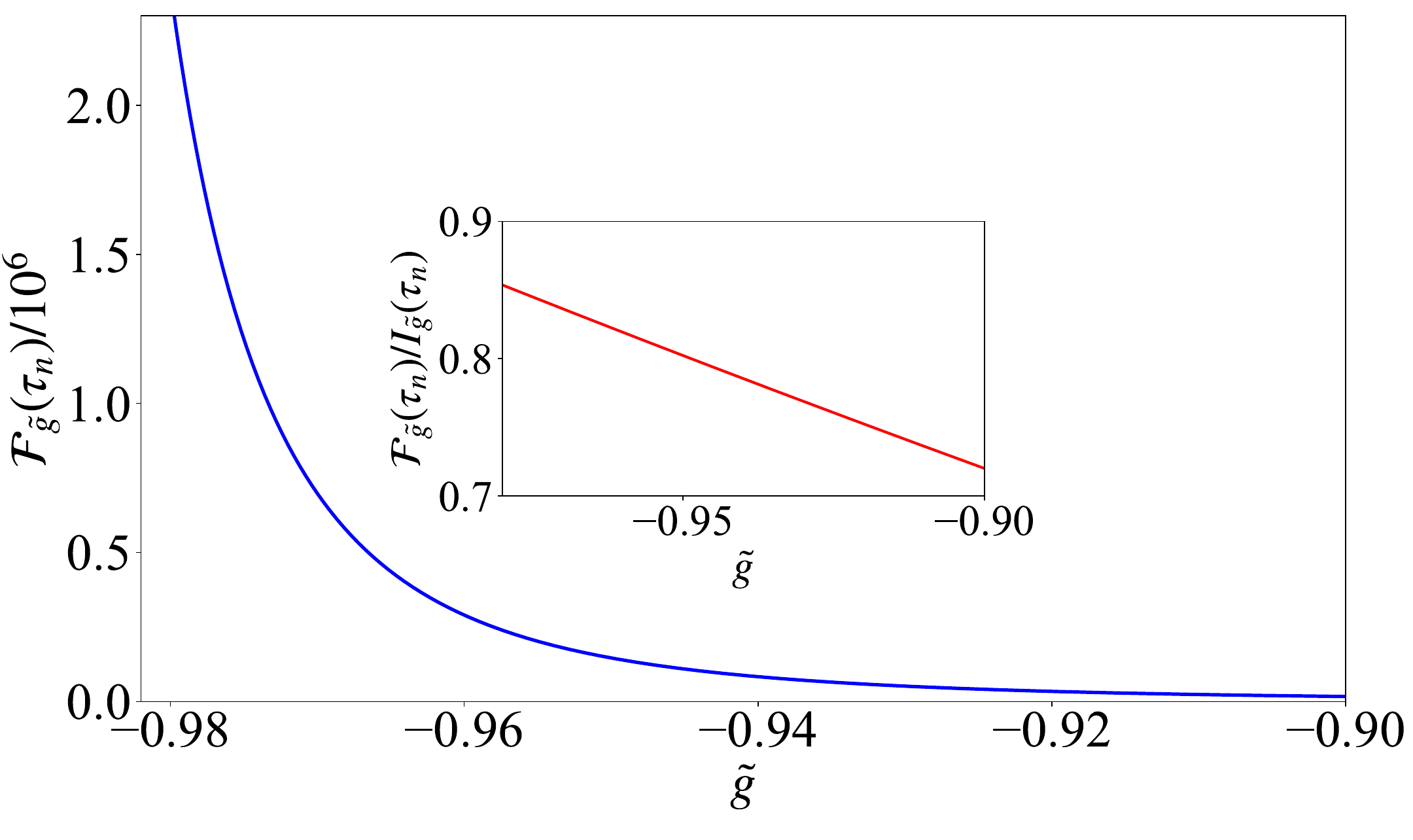}
\caption{The peak of the inverted variance $\mathcal{F}_{\tilde{g}}(t)$ of the CIM as a function of the parameter ${\tilde{g}}$ for an evolution time $\tau_n$ with $\alpha=1$. The inset shows that the local maxima of the inverted variance $\mathcal{F}_{\tilde{g}}(\tau_n)$ reaches  the same order of $I_{\tilde{g}}(\tau_n)$. }
\label{FIG:CIMIgFg}
\end{figure}
Finally, Fig.~\ref{FIG:CIMIgFg} shows the peak of the inverted variance $\mathcal{F}_{\tilde{g}}(\tau_n)$ as a function of $\tilde{g}$. Its behavior is completely consistent with the LMG model shown in Fig.~\ref{FIG:LMG} (d), exhibiting strong divergence as $|\tilde{g}|\rightarrow 1$. The inset confirms that $\mathcal{F}_{\tilde{g}}(\tau_n)$ reaches the same order of magnitude as the QFI $I_{\tilde{g}}(\tau_n)$, demonstrating the optimality of the quadrature measurement on the CIM platform. The strong agreement between the CIM and LMG model results demonstrates that the CIM effectively simulates both the QPT and the criticality-enhanced sensing characteristics of the LMG model via the mapped effective Hamiltonian.

\section{Conclusions}\label{sec:Conclusions}

We have theoretically proposed and validated an effective scheme for simulating the  LMG model based on the CIM. The scheme's core is the mapping of the spin variables onto the phase states of DOPOs and the realization of all-to-all interactions through tunable optical mutual injection. Through detailed theoretical derivation, we have demonstrated that the effective Hamiltonian of the CIM becomes equivalent to the LMG model after unitary transformation and pseudo-spin mapping, with the parameter correspondence $\lambda = -J/2$ and $h = \Omega/2$. Based on this, we have analyzed the ground-state energy and its derivatives, clearly revealing a second-order QPT at $\Omega_c = \pm J$ with non-analytic behavior consistent with the LMG model.

Beyond simulating the phase transition, we have further explored the application of the LMG model and the CIM platform to quantum metrology. We show that by monitoring the dynamics of the quadrature $P$, its susceptibility $\chi_g$ to the coupling parameter diverges near the critical point. The QFI and the inverted variance $\mathcal{F}_g(t)$ both exhibit criticality-enhanced behavior, indicating the potential for achieving measurement precision beyond the standard quantum limit. 

Our work successfully extends the application of the CIM from combinatorial-optimization questions to the fields of quantum many-body simulation and criticality-enhanced sensing. The programmability, exceptional scalability, and all-to-all connectivity of the CIM make it a powerful and flexible platform for these studies. Our work bridges the fields of quantum optics and condensed matter physics by providing a photonic platform to explore fundamental quantum magnetic models. Furthermore, the platform’s compatibility with real-time measurement and control opens avenues for simulating non-equilibrium quantum dynamics, such as quench and ramp protocols across critical points. Future work will focus on the experimental realization of this scheme and the investigation of non-equilibrium critical dynamics for sensing within the CIM framework.

\section*{acknowledgments}

This work is supported by the National Natural Science Foundation of China under Grant No.~62461160263, Innovation Program for Quantum Science and Technology under Grant No.~2023ZD0300200,  and
Guangdong Provincial Quantum Science Strategic Initiative under Grant No.~GDZX2505004.

{
\appendix
\section{The validity of the adiabatic elimination and the role of the quantum noise terms}\label{appendixA}

Our elimination procedure follows the standard Heisenberg-Langevin treatment widely used for the DOPO-based CIMs \cite{Inui2020pra}. We start from Eq.~\eqref{eq:Hall} to obtain the Heisenberg-Langevin equations for the $N$ DOPO pulses, i.e.,
\begin{align}
\frac{d a_{\text{p}j}}{dt} =& - \gamma_{\text{p}} a_{\text{p}j} + \varepsilon - \frac{\kappa}{2} a_{\text{s}j}^2+\sqrt{\gamma_p}\xi_1, \label{eq:HL_pump} \\
\frac{d a_{\text{s}j}}{dt} =& -(i\Delta + \gamma_{\text{s}}) a_{\text{s}j} + \kappa a_{\text{s}j}^{\dagger}a_{\text{p}j} + \zeta \sum_{k \neq j} (a_{\text{\rm c}jk} - i a_{\text{\rm c}kj})\nonumber \\
&+\sqrt{\gamma_s}\xi_2 \label{eq:HL_signal}\\
\frac{d a_{\text{c}jk}}{dt} =& - \gamma_{{\rm c}} a_{{\rm c}jk} - \zeta a_{\text{s}j} - i\zeta a_{\text{s}k}+\sqrt{\gamma_c}\xi_3,
\label{eq:HL_couppling}
\end{align}
where the quantum-noise operators satisfy $\langle \xi_i^\dagger(t)\xi_i^\dagger(t')\rangle = 0$ and $\langle \xi_i(t)\xi_j^\dagger(t')\rangle = 2\,\delta_{ij}\delta(t-t')$. Noting that Eq.~\eqref{eq7.1} in the main text is obtained under ideal conditions. Without $\Delta$, it would have no effect on the adiabatic elimination of the pump mode. The prefactor $\sqrt{\gamma_{p/s/c}}$ in Eqs.~\eqref{eq:HL_pump}-\eqref{eq:HL_couppling} is the standard Markovian input-output quantum Langevin normalization, i.e., it guarantees that dissipation at rate $\gamma$ is accompanied by vacuum fluctuations whose $\delta$-correlated strength preserves equal-time commutation relations and yields the correct steady-state fluctuation level. 

The formal solution for the pump operator in Eq.~\eqref{eq:HL_pump} is
\begin{align}
a_{{\rm p} j}(t) =& e^{-\gamma_p t}a_{{\rm p} j}(0) + \int_0^t d\tau\, e^{-\gamma_p(t-\tau)}
\Big[\epsilon - \frac{\kappa}{2}a_{{\rm s} j}^2(\tau)\Big] \nonumber \\
&+ \sqrt{\gamma_p}\int_0^t d\tau\, e^{-\gamma_p(t-\tau)}\xi_1(\tau).
\end{align}

The validity of the adiabatic elimination relies on the hierarchy of time scales, i.e., $\gamma_p \gg \gamma_s$. Under this condition, the pump mode relaxes much faster than the signal mode, i.e., $1/\gamma_p \ll 1/\gamma_s$, allowing us to approximate $a_s(\tau) \approx a_s(t)$ within the integral kernel.  
The adiabatic solution is then obtained as
\begin{equation}
a_{{\rm p} j}(t) \simeq \frac{\epsilon}{\gamma_p}-\frac{\kappa}{2\gamma_p}a_{{\rm s} j}^2
+ \frac{1}{\sqrt{\gamma_p}}\xi_1(t)
+ \mathcal{O}\!\left(\gamma_p^{-2}\right).
\label{eq:ap_adiabatic}
\end{equation}
This reproduces the expression after the adiabatic elimination, including the pump-noise contribution.

A similar time-scale separation argument applies to the adiabatic elimination of the coupling fields $a_{cjk}$s, i.e., 
\begin{align}
a_{{\rm c} jk} &= -\frac{\zeta_{jk} }{\gamma_{c}}\left( a_{{\rm s} j} + ia_{{\rm s} k} \right)+\frac{1}{\sqrt{\gamma_c}}\xi_3(t), \label{eq:Aa_{cjk}} \\
a_{{\rm c} kj} &= -\frac{\zeta_{kj} }{\gamma_{c}}\left( a_{{\rm s} j} + ia_{{\rm s} k} \right)+\frac{1}{\sqrt{\gamma_c}}\xi_3(t). \label{eq:Aa_{ckj}}
\end{align}
When the coupling modes relax on a much faster time scale than the signal modes, typically $\gamma_c\gg\gamma_s$, it follows the signal quasi-instantaneously and yields Eq.~\eqref{eq:HC} as the leading-order slaving relation. If $\gamma_c$ is not the fastest rate, retardation effects and additional quantum-noise terms may become non-negligible, in which case one should retain the coupling dynamics explicitly {\color{blue}\cite{Inui2020pra,Zhou2021pra}}.

Substituting Eq.~\eqref{eq:ap_adiabatic}-\eqref{eq:Aa_{ckj}} into Eq.~\eqref{eq:HL_signal} gives the simplified stochastic differential equation (SDE) for the signal mode, i.e.,
\begin{align}
\frac{d a_{{\rm s}j}}{dt}
\simeq &-(\gamma_s+i\Delta) a_{{\rm s}j}
\!\!+ \!\!\frac{\kappa\epsilon}{\gamma_p}a_{{\rm s}j}^\dagger
\!\!- \!\!\frac{\kappa^2}{2\gamma_p}a_{{\rm s}j}^\dagger a_{{\rm s}j}^2\!\!-\!\!\frac{2}{\gamma_c}\sum_{k\neq j}\zeta_{jk}^2 a_{{\rm s}j}  \nonumber \\
&
+\sqrt{\gamma_s}\,\xi_2(t)+ \frac{\kappa}{\sqrt{\gamma_p}}\,a_{{\rm s}j}^\dagger\,\xi_1(t)+\frac{1}{\sqrt{\gamma_c}}\sum_{k\neq j}[\zeta_{jk} \nonumber \\
&(\xi_3(t)-i\xi_3(t))].
\label{eq:reduced_SDE_keep}
\end{align}
{The noise terms for the pump and coupling fields which have been discarded will now appear here as multiplicative noise terms. The simplification of the noise term of the pump field is justified by considering its magnitude relative to the signal noise. From Eq.~\eqref{eq:reduced_SDE_keep}, the signal noise enters as $\sqrt{\gamma_s}\,\xi_2(t)$, while the pump-noise term enters as $(\kappa/\sqrt{\gamma_p})\,a_{{\rm s}j}^\dagger\,\xi_1(t)$. The ratio $R$ of the contribution of the pump noise to that of the signal vacuum noise scales approximately as
\begin{equation}
R \sim \frac{\kappa \langle |a_s| \rangle / \sqrt{\gamma_p}}{\sqrt{\gamma_s}} \propto \sqrt{\frac{\gamma_s}{\gamma_p}} \sqrt{\langle n_s \rangle_{\rm sat}},
\end{equation}
where we assume the system is in the saturation regime. In our CIM implementation, since $\gamma_p \gg \gamma_s$, the contribution of the pump noise by $1/\sqrt{\gamma_p}$ is strongly suppressed and thus $R \ll 1$ in the parameter regime of interest.}

To directly quantify the impact of the adiabatic elimination of the pump noise, we compare the photon-number dynamics of the signal
$\langle n_s(t)\rangle\equiv \langle a_s^\dagger a_s\rangle$ obtained from three stochastic descriptions. In Fig.~\ref{FIG:SM}, the blue solid line is plotted from the full Heisenberg-Langevin equations with the both noises retained, i.e., Eq.~\eqref{eq7.1}-\eqref{eq7} in the main text. The orange dashed line is plotted from the simplified signal-mode SDE, i.e., Eq.~\eqref{eq:reduced_SDE_keep}, where the pump and coupling modes are adiabatically eliminated but the pump noise and coupling noises are kept. The green dotted uses the simplified SDE without the pump noise and the coupling noise. In this way, we can justify the adiabatic elimination. \\

\begin{figure}[t]
\centering
\includegraphics[width=0.47\textwidth,trim=0 0 0 0,clip]{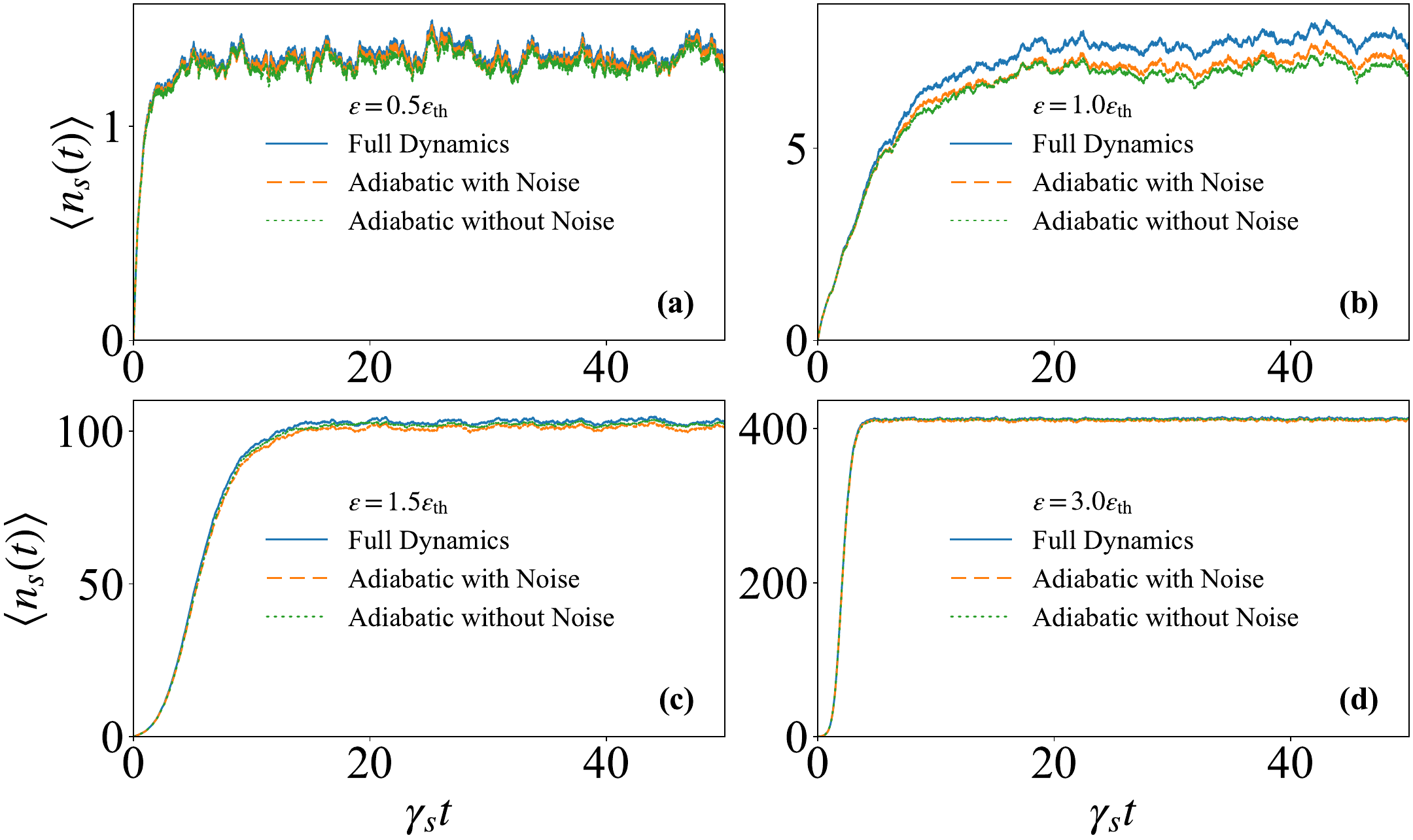}
\caption{{Signal photon-number dynamics $\langle n_s(t)\rangle$ for different pump strengths, comparing the full model (blue solid line) and simplified signal-mode models with (without) the pump noise and the coupling noises (orange dashed /green dotted line). (a)–(d) correspond to $\epsilon/\epsilon_{\rm th}=0.5,\,1.0,\,1.5,$ and $3.0$, respectively. Our parameters are $\gamma_s=0.1, \gamma_p=200\gamma_s, \gamma_c=300\gamma_s$, and $\kappa=1.4\gamma_s$. In addition, we consider the system with a finite size, i.e., the number of the DOPO pulses is $N=3000$.}}
\label{FIG:SM}
\end{figure}

Crucially, our CIM-to-LMG mapping is formulated in the large $N$ regime, and targets the above-threshold operating condition of CIM, where each DOPO pulse is typically operated above threshold and evolves into a macroscopic phase-bistable coherent state. In this above-threshold saturation limit, the photon number quickly achieves a large value and the dynamics is predominantly governed by deterministic gain-saturation drift, while the quantum noises produce only relatively-small fluctuations around the mean trajectory. Consistent with this physical picture, Fig.~\ref{FIG:SM} shows that different cases, indicating that the residual contributions of the pump noise and the coupling noise neglected in the effective-Hamiltonian approach is practically negligible under $\gamma_p\gg\gamma_s$ and $\gamma_c\gg\gamma_s$ for the parameter regime considered here. As the system is operated further above threshold, the difference between keeping vs neglecting the pump noise and the coupling noise becomes even less visible. On the contrary, very close to threshold or without a clear time-scale separation, e.g., $\gamma_p \sim \gamma_s$ and $\gamma_c \sim \gamma_s$, one should stick to the full stochastic dynamics, and the  effective-Hamiltonian description may fail. Furthermore, when we consider a system with a finite size, e.g. $N=3000$, Fig.~\ref{FIG:SM} further shows that once the pump strength is far above threshold, the photon number of  the signal mode $\langle n_s\rangle$ can become so large that it even exceeds $N$, i.e., $\langle n_s(t)\rangle > N$. In this regime, our low-excitation approximation is still valid. We notice that in Ref.~\cite{SA100000}, about $10^5$ qubits has been demonstrated.}

\bibliography{MSQbib}
\end{document}